\documentclass[manuscript]{aastex}

\newcommand{\rearth}{R$_{\oplus}$}

\newcommand{\msun}{M$_{\odot}$}

\newcommand{\mps}{m~s$^{-1}$}

\slugcomment{Submitted to the Astrophysical Journal}

\shorttitle{Small Planets around the Coolest {\it Kepler} Stars}
\shortauthors{Gaidos et al.}

\begin{document}


\title{On the Nature of Small Planets around the Coolest {\it Kepler} Stars \footnotemark}

\footnotetext[1]{Some data were obtained at the W.~M. Keck
  Observatory, which is operated by the California Institute of
  Technology, the University of California and NASA, and made possible
  by the financial support of the W.~M. Keck Foundation.}

\author{Eric Gaidos\altaffilmark{2}} 
\affil{Department of Geology and Geophysics, University of Hawai`i at M\={a}noa, Honolulu, HI 96822}

\email{gaidos@hawaii.edu}

\author{Debra A. Fischer} 
\affil{Department of Astronomy, Yale University, New Haven, CT 06520}

\author{Andrew W. Mann}
\affil{Institute for Astronomy, University of Hawai`i at M\={a}noa, Honolulu, HI 96822}

\and

\author{S\'{e}bastien L\'{e}pine} 
\affil{Department of Astrophysics, American Museum of Natural History, New York, NY 10024}

\altaffiltext{2}{Visiting Professor, Pufendorf Institute, Lund University, Sweden}

\begin{abstract}
  We constrain the densities of Earth- to Neptune-size planets around
  very cool ($T_e$=3660-4660~K) {\it Kepler} stars by comparing 1202
  Keck/HIRES radial velocity measurements of 150 nearby stars to a
  model based on {\it Kepler} candidate planet radii and a power-law
  mass-radius relation.  Our analysis is based on the presumption that
  the planet populations around the two sets of stars are the same.
  The model can reproduce the observed distribution of radial velocity
  variation over a range of parameter values, but, for the expected
  level of Doppler systematic error, the highest Kolmogorov-Smirnov
  probabilities occur for a power-law index $\alpha \approx 4$,
  indicating that rocky-metal planets dominate the planet population
  in this size range.  A single population of gas-rich, low-density
  planets with $\alpha = 2$ is ruled out unless our Doppler errors are
  $\ge$5~\mps, i.e., much larger than expected based on observations
  and stellar chromospheric emission.  If small planets are a mix of
  $\gamma$ rocky planets ($\alpha = 3.85$) and 1-$\gamma$ gas-rich
  planets ($\alpha = 2$), then $\gamma > 0.5$ unless Doppler errors
  are $\ge$4~\mps.  Our comparison also suggests that {\it Kepler}'s
  detection efficiency relative to ideal calculations is less than
  unity.  One possible source of incompleteness is target stars that
  are misclassified subgiants or giants, for which the transits of
  small planets would be impossible to detect.  Our results are robust
  to systematic effects, and plausible errors in the estimated radii
  of {\it Kepler} stars have only moderate impact.
\end{abstract}


\section{Introduction}

The discovery of planets around other stars has placed our Solar
System in context and stimulated speculation on the frequency of
habitable planets and life in the Universe.  Very cool dwarf stars
(with late K and early M spectral types) are of special significance
to such investigations because the two principle detection techniques,
Doppler radial velocity (RV) and transit photometry, are more
sensitive to smaller planets around smaller stars.  Such stars are
also much less luminous than solar-type stars, the circumstellar
habitable zone is closer \citep{Kasting1993}, and planets within the
habitable zone are therefore more detectable \citep{Gaidos2007}.
These stars test models of planet formation: for example,
core-accretion models predict fewer gas giants and more "failed" cores
\citep{Laughlin2004,Kennedy2008}, consistent with the lower frequency
of giant planets and higher frequency of low-mass planets compared to
G stars \citep{Johnson2007a,Cumming2008,Mayor2009}.  Finally, late K
and early M dwarfs constitute three-quarters of all stars in the
Galaxy, and their contribution weighs heavily in any cosmic accounting
of planets or life.

Most confirmed exoplanets have been found by the Doppler technique,
which can detect planets of a few Earth masses on short-period orbits
around bright late F- to early K-type stars
\citep{Mayor2009,Howard2010}.  There are also Doppler searches for
planets around very cool dwarfs
\citep{Zechmeister2009,Apps2010,Bean2010,Forveille2011}.  The CoRoT
and {\it Kepler} missions have successfully extended the search for
small planets to space using the transit technique.  The {\it Kepler}
spacecraft is monitoring $\sim$150,000 stars, including approximately
24,000 K-type stars and 3000 M-type stars \citep{Batalha2010}, and has
discovered hundreds of candidate planets with radii $R_p$ as small as
$\sim$0.8~\rearth~\citep{Borucki}.  The distribution with $R_p$ peaks
near 2\rearth~and at the completeness limit of {\it Kepler}
\citep{Howard2011}.

\begin{figure}
\epsscale{0.7}
\plotone{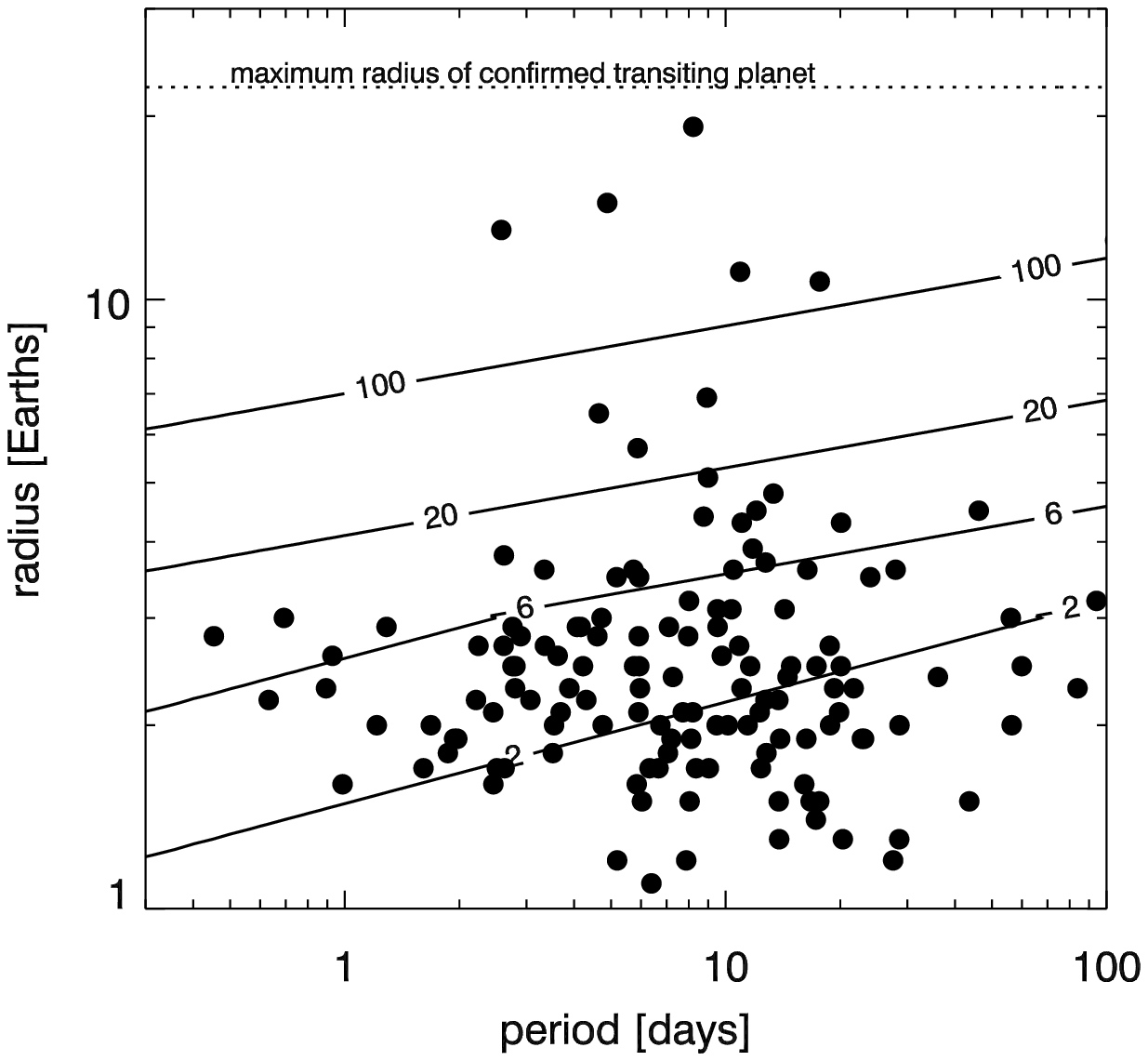}
\caption{{\it Kepler} candidates plotted by orbital period and planet
  radius $R_p$, and contours of constant radial velocity variation
  (RMS = 2, 6, 20, 100~\mps) assuming a circular orbit, orbital
  inclination of 60$^{\circ}$, and a mass given either by an average
  of confirmed transiting planets with similar radius (if $R_p
  >$3\rearth) or proportional to radius squared (if $R_p <$3\rearth).
  If the small planets are rocky then mass will be higher
  (proportional to $R_p^4$) and the RV RMS contours will be lower.
  Although many {\it Kepler} planets would be very difficult to
  individually detect (RMS $< 6$~\mps), they will aggregately
  contribute to significant RV variation, especially if they are
  composed primarily of rock and metal. \label{fig.koi}}
\end{figure}

In principle, the mass $M_p$ of a transiting planet can be uniquely
determined by Doppler observations and mass and radii compared with
theoretical relationships.  The mean density of scores of giant
planets and a handful of objects between the size of Earth and Neptune
orbiting nearby stars have been determined in this manner
\citep{Gillon2007,Charbonneau2009a,Hartman2011,Winn2011,Demory2011}.
This technique has also successfully confirmed candidate planets
around the brightest CoRoT and {\it Kepler} stars, including two with
masses only a few times that of Earth \citep{Batalha2011,Hatzes2011}.
Comparison with a mass-radius relationship (MRR) can discriminate
between denser planets composed of silicates and metal
(``super-Earths''), and less dense planets with substantial envelopes
of ices and hydrogen-helium gas (``ocean planets'' or
``mini-Neptunes'') \citep{Seager2007}.  However, the Doppler signal
expected from many {\it Kepler} candidate planets is comparable to
total instrument noise and stellar ``jitter'' (2-3~\mps, Figure
\ref{fig.koi}).  RV measurements can be ``phased'' to the
transit-determined orbit, achieving greater sensitivity.
Unfortunately, the great majority of cool {\it Kepler} stars are too
faint ($K_p > 13$) to achieve the required high SNR even using 10-m
telescopes.

Instead, Doppler observations of a sample of {\it nearby}, brighter
stars can constrain the masses and mean densities of planets around
corresponding {\it Kepler} stars, assuming both samples host the same
planet population.  Every planet will contribute to RV variance and
the aggregate effect in excess of instrument errors and the noise from
the stellar atmosphere (``jitter'') can be detected.  Given {\it
  Kepler}-determined orbits and planet radii and a hypothetical MRR,
the cumulative distribution of RV variation can be predicted and
compared to that from the nearby population.  For a given distribution
of observed radii, denser, rocky planets will generate greater RV
variation, while less dense, ice- or gas-rich planets will produce
smaller variation.

This approach exploits both the orbital information from {\it Kepler}
and the collective RV signal from the entire population.  As with RV
follow-up of individual transiting planets, planets are first detected
by transit ({\it Kepler}), then characterized by Doppler observations.
{\it Kepler} observations would provide an exact description of a
equivalent nearby population only in the limit of an infinite sample,
and thus the finite size of the candidate planet sample introduces
uncertainty.  We show that this uncertainty is not debilitating.  This
method also rests on two assumptions: (i) the planet populations of
the Doppler and {\it Kepler} samples are statistically the same, and
(ii) the MRR of small planets can be described by simple empirical
relations.  We discuss the validity of both of these assumptions.

We carry out such a combined transit-Doppler analysis, predicting the
statistical distribution of RV variation in the M2K survey of late K
and early M dwarfs \citep{Apps2010}.  We use the {\it Kepler}
distribution of candidate planet radii, corrected for detection
efficiency, and assume a single parametric MRR.  We compare the
predicted and observed distributions to constrain the MRR and hence
the compositions of the small planets these stars host.

\section{Data \label{sec.data}}

{\it Doppler survey:} The M2K survey has obtained 1406 RV measurements
of 172 late K and early M dwarfs, with at least 3 measurements for
each star.  Stars were selected from the SUPERBLINK proper motion
catalog \citep{Lepine2005} based on $V$-$J$ color and parallax- or
proper-motion-based absolute magnitudes \citep{Lepine2011}, and
confirmed by moderate-resolution spectroscopy.  We excluded active
stars with detectable emission in H$\alpha$ or in the 90th percentile
of emission in the HK lines of Ca II, and another 6 stars with
problematic template spectra.  The remaining stars are not
exceptionally active, with median $R'_{HK} = -4.70$ and the vast
majority have $-5 < R'_{HK} < -4.5$ (see inset in Figure
\ref{fig.activity}).  For stars with $B-V \approx 1$ these activity
levels correspond to ages of 1-10~Gyr \citep{Mamajek2008}.  Targets
have apparent magnitudes of $V=8-12$; most have $V=9-10$.

Doppler spectra are obtained with the red channel of the HIRES
spectrograph on the Keck I telescope \citep{Vogt1994}.  Exposure times
are adjusted to achieve SNR = 200.  Absorption lines of molecular
iodine are used as a rest-frame reference against which to measure the
Doppler shift of features in the stellar spectrum.  The shift is
determined by minimizing the difference between the spectrum and a
model combining an observed spectrum of the star without iodine and
one of iodine imposed on the featureless spectrum of a B star
\citep{Marcy1992,Butler1996}.  The error-weighted mean is subtracted
from the measurements of each star and the RMS is calculated (Table
\ref{tab.obs}).

The effective temperature $T_e$ of each star is estimated from the
$V$-$K$ color and an empirical relation
\begin{equation}
\log T_e \approx 3.9653 - 0.164 ({\rm V}-{\rm K}) + 0.0168 ({\rm V}-{\rm K})^2,
\end{equation}
which has an accuracy of 1\% \citep{Benedetto1998}.  We estimate
stellar mass $M_*$ using an empirical relation $\log \left( M_*/
  M_{\odot} \right) = 1.5 \log \left(T_e/5780 \right) + 0.02$ based on
a Yale-Yonsei 5 Gyr isochrone \citep{Demarque2004}.  The metallicites
of 95 stars have been estimated using the Spectroscopy Made Easy code
\citep{Valenti1996}.  The standard deviation of [Fe/H] in our sample
is $\pm$0.21 dex, and the concomitant error in stellar mass due to the
use of a solar-metallicity isochrone is $\sim$0.02\msun, which we
ignore.

{\it Kepler targets and planets:} We use the Quarter 2 {\it Kepler}
target list from the Multimission Archive (STScI).  {\it Kepler}
candidate planets are taken from \citet{Borucki}, who report $R_p$
based on stellar radius $R_*$, the orbital period $P$, and the
estimated $T_e$ and surface gravity $\log g$ of the host star.
Stellar parameters are based on the multi-passband photometry and
Bayesian analysis of the Kepler Input Catalog (KIC) \citet{Brown2011}.
We consider only putative dwarf stars with $4 < \log g < 4.9$.

{\it Effective temperature range:} We choose a $T_e$ range that
includes a substantial number of stars from each sample and maximizes
the similarity in the temperature distributions as assayed by the
Kolmogorov-Smirnov (K-S) statistic.  For an interval of 1000~K, that
range is 3660-4660~K (K-S probability $= 4.7 \times 10^{-3}$).  This
includes 150 M2K stars (1202 measurements) and 10,018 {\it Kepler}
target stars, the latter having 138 candidate planets, and excludes
the 410 very coolest {\it Kepler} target stars and 6 hotter M2K stars.
The mean effective temperatures of the M2K and {\it Kepler} subsamples
are 4230~K and 4200~K, respectively.  The low K-S probablity reflects
the narrower distribution of M2K stars within this range of $T_e$
compared to the {\it Kepler} sample (Figure \ref{fig.teff}).  We
speculate on the possible impact of this difference on our analysis in
Section \ref{sec.discussion}.

\section{Model \label{sec.model}}

{\it Planet frequency:} The expected frequency of the $i$th planet
candidate in the {\it Kepler} survey is $1/s_i$, where $s_i = \Sigma_j
p_{ij}q_{ij}$, $p_{ij}$ is the geometric probability of a transiting
orbit around the $j$th star, and $q_{ij}$ is the probability of
detection if the planet is on a transiting orbit.  $s_i$ is the
expected number of stars around which a planet would be detected, if
every star had this planet on its particular orbit.  For example, a
planet that could have been detected around 100 stars, but has been
found once, has a most likely occurence rate of 1\%.  For planets that
are small compared to their host stars and on nearly circular orbits,
the transit probability is:
\begin{equation}
p = 0.238FP^{-2/3}M_*^{-1/3}R_*,
\end{equation}
where $P$ is in days and $F = T/P$ if $P > T$, where $T$ is the
observation period (120~d), or else $F = 1$.  $M_*$ and $R_*$ are in
solar units.  A planet is detected if $SNR = \delta/\sigma \ge 7$
\citep{Borucki}, where $\delta$ is the transit depth and $\sigma$ is
the noise over the entire transit.  In our Monte Carlo calculations
(see below) $q_{ij}$ only takes on values of 0 or 1 depending on
whether $SNR \lessgtr 7$.  Assuming uncorrelated noise,
\begin{equation}
SNR = \frac{\delta}{\sigma_{30}}\sqrt{\frac{N\Delta}{30}},
\end{equation}
where $\sigma_{30}$ is the noise per 30-minute integration, $N$ is the
number of observed transits, and $\Delta$ is the transit duration in
minutes.  The transit depth is $\delta \approx 8.4 \times 10^{-5}
(R_p/R_*)^2$, where $R_p$ is in Earth units.  The noise per 30-min
integration as a function of {\it Kepler} magnitude $K_P$ is
$\sigma_{30} \approx 10^{(K_P-13)/5 - 4}$ \citep{Koch}.  We multiply
this by a factor drawn randomly from the distribution in Figure 4 of
\citet{Koch} to account for stellar variability.  Stars with factors
$>10$ are assigned a factor of 10.  The number of observed transits is
the largest integer less than $T/P$.  (Three cases where $P > T$ and
$N=1$ were confirmed by the {\it Kepler} team using later
observations.)  The transit duration for a circular orbit, averaged
over all possible impact parameters, is
\begin{equation}
  \tau \approx 85 R_* M_*^{-1/3}P^{1/3} {\rm min}.
\end{equation}
To account for incompleteness or overestimation of the detection
efficiency of {\it Kepler}, we multiply $s_i$ by a constant parameter
$C$, where $0 < C \le 1$.  We use a single, uniform value for
detection efficiency both as a necessary simplification and because it
can describe one possible cause of detection inefficiency - the
presence of giant stars in the target list (Section
\ref{sec.results}).  We do not correct for false positives, probably
5-10\% \citep{Morton2011}. $C > 1$ is possible but unlikely if the
false-positive rate is low, and we do not consider values of $C$
$<0.2$.

{\it Mass-radius relations:} For the MRR of planets with $R_p >
3$\rearth, i.e. Neptune size or larger, we use the masses and radii of
120 confirmed transiting planets \citep{Schneider2011}. $M_p$ is
calculated using the mean density of the 8 such planets with radii
closest to that of the {\it Kepler} object.  Smaller planets with
radii $\le3$\rearth~are described by a single population with $M_p =
R_p^{\alpha}$ (Earth units).  Although the MRRs of solid planets
(rock/ice/metal) are not expected to precisely follow power laws
\citep{Fortney2007a,Seager2007}, a power law with $\alpha \approx
3.85$ is a reasonable approximation for a planet with an Earth-like
ratio of silicates to metal, and little gas.  If planets have acquired
and retained a substantial H-He envelope, and the mass fraction of the
envelope increases with $M_p$, we expect $\alpha \ll 4$.  For example,
$M_p \sim R_p^2$ describes a continuum between Earth and Neptune, and
gas-rich super-Earths may have $\alpha \le 0$ \citep{Rogers2011}.  Of
course, the small planets may be a mix of both rocky- and gas-rich
objects and we entertain this scenario in Section \ref{sec.results}.

{\it Radial velocity errors:} Appropriate modeling of RV errors is
crucial to this analysis.  The median standard deviation of formal
(including Poisson) errors is 1.3~\mps, and we use the actual formal
errors in our calculations.  Fifteen pairs of RV measurements taken
within 6~hr show additional total systematic error of $\approx
3$~\mps. We assume that additional systematic instrument errors and
stellar noise (``jitter'') are uncorrelated between observations and
gaussian-distributed, but we examine the effect of correlated
instrument errors in Section \ref{sec.tests}.  For instrument noise we
use a fixed RMS of 1.6~\mps~based on observations showing this to be
the ``basement'' level of systematic noise among a large number of
HIRES observations of K stars \citep{Isaacson2010a}.  Stars do not
exhibit a monotonic level of jitter.  Figure \ref{fig.activity} shows
the distribution of total systematic noise (instrument plus stellar
jitter) predicted for 100 M2K stars based on their Ca II HK emission,
$B$-$V$ colors, and the equations in \citet{Isaacson2010a}.  We adopt
a Rayleigh formula for the distribution of the jitter RMS $\sigma_*$
among all stars in the sample,
\begin{equation}
\label{eqn.rayleigh}
p(\sigma_*) = \frac{\sigma_*}{\sigma_0} \exp \left(\frac{-\sigma_*}{\sigma_0}\right),
\end{equation}
where we term $\sigma_0$ the {\it magnitude} of the jitter.  (In
Section \ref{sec.tests} we also try an exponential distribution.)  The
RMS jitter in an {\it ensemble} of stars with a Rayleigh distribution
is $\sqrt{2}\sigma_0$.  Our 6~hr systematic noise level of 3~\mps~can
be explained if $\sigma_0 = 1.8$~\mps.  The predicted jitter
distribution is best described by $\sigma_0 = 1.7\pm0.1$~\mps~(Figure
\ref{fig.activity}), consistent with our observations of 3 \mps~total
RMS .  Additional noise due to stellar rotation and starspots may
occur on longer timescales \citep{Barnes2011}, and we perform
calculations with $\sigma_0$ over the range 1.5-4.5~\mps.  However, we
consider values near the upper limit, corresponding to an average
systematic noise of 6.5~\mps, highly implausible because of the
absence of active stars in in our sample (inset of Figure
\ref{fig.activity}).  This is discussed further in Section
\ref{sec.discussion}.

{\it Radial velocity calculations:} We predict the distribution of RV
RMS for each set of parameter values by generating 10,000 Monte Carlo
systems, with host stars selected with replacement from the M2K
survey, and orbital inclinations drawn from an isotropic distribution.
Each {\it Kepler} candidate planet has a probability $1/s_i$ of being
added to each star.  This ignores any autocorrelation between the
presence of planets.  Masses are assigned to each planet using the
{\it Kepler} radius and the MRR.  We ignore all planet candidates with
radii larger than the largest confirmed transiting planet ($\sim$2
Jupiter radii) as main sequence companions or false positives.  The RV
variation induced by each planet is calculated from the planet mass,
host star mass, and system inclination.  Orbits are assumed to be
approximately coplanar \citep{Lissauer2011b}.  Radial velocities are
calculated using the actual epochs of observations and random mean
anomalies at the first epoch.  We draw longitudes of perihelion from a
uniform distribution and orbital eccentricities from a Rayleigh
distribution with mean of 0.225 \citep{Moorhead2011}.  We add formal
and systematic errors to the simulated radial velocities, subtract the
error-weighted mean, and calculate the RMS.  To filter binary stars,
we remove observed and predicted systems whose RMS exceeds a specified
cutoff $B$.

{\it Statistical comparison:} The model and observed distributions are
compared using the two-sided Kolmogorov-Smirnov (K-S) test and the
two-sample Kuiper test; the latter is sensitive to the tails of a
distribution as opposed to the median.  The four parameters of the
model are the MRR parameter $\alpha$, jitter magnitude $\sigma_0$,
binary cutoff $B$, and completeness $C$.  In Section \ref{sec.results}
we introduce a fifth parameter $\gamma$ that describes a mixed
population of rocky and Neptune-like planets.

\section{Results \label{sec.results}}

A byproduct of our analysis is an estimate of the average number of
planets per star: $\sum_i s_i^{-1}$.  We find that 30\% of {\it
  Kepler} stars with $T_e$ = 3660-4660~K have planets with $R_e >
2$\rearth~and $P < 50$~d.  This is in agreement with the findings of
of \citet{Howard2011}.  The frequency of giant planets ($R_p >
0.8R_J$) in our sample is 2.4\%, close to that estimated in Doppler
surveys \citep{Johnson2010}.  This indicates minimal bias in our Monte
Carlo reconstruction of the discrete {\it Kepler} sample because any
effect should be most pronounced for the rarest (largest) planets.

The observed cumulative distribution of RV RMS (points in Figure
\ref{fig.rmsdist}) has an accelerating rise below 3~\mps~from gaussian
noise, a logarithmic increase over 3-10~\mps~from the combined effect
of systematic error and planets not resolved by Doppler observations,
and a tail beyond 10~\mps~from giant planets and low-inclination
binary stars.  The best-fit models (e.g., solid line) agree with the
observed distribution with a K-S probability $>$90\%.  The K-S and
Kuiper statistics are largely congruent and hereafter we show only the
former.  95\% confidence intervals in the uncertainty due to the
finite size of the {\it Kepler} planet sample were calculated using
200 bootstrap-resampled planet populations and are plotted as dashed
lines in Figure \ref{fig.rmsdist}.  (These illustrate deviations from
the best-fit cumulative distribution, and are not cumulative
distributions themselves, which can never reverse).  The high RMS tail
of the distribution contains few systems and is most poorly
reproduced.

Jitter magnitude, completeness, and MRR parameter $\alpha$ influence
the predicted distribution of RV variation in similar ways, and
different combinations of parameter values can reproduce the
observations.  Values that produce high K-S probabilities describe a
locus in $C-\alpha-\sigma_0$ space.  In contrast, our results are
insensitive to the binary cutoff $B$ for reasonable values; $B =
110$~\mps~is used in all analyses.  This value excludes 24 systems and
implies that $\sim70$\% of M2K stars are single. This is an upper
limit because M2K excludes known spectroscopic and close ($<2$
arc-sec) binaries, and is not sensitive to wide (but unresolved)
binaries.  This fraction is intermediate the single star fraction of
40\% for G stars and 80\% for M stars \citep{Lada2006}.

We first performed calculations assuming $C = 1$ and allowing $\alpha$
to vary from 2 to 5, and $\sigma_0$ to vary from 1.5 to 4.5~\mps.
This range of $\sigma_0$ is intended to capture the locus of high K-S
probabilities over the entire plausible range of $\alpha$; high values
of $\sigma_0$ clearly contravene our observations and predictions
based on chromospheric emission (Figure \ref{fig.activity}).  Contours
of 0.01, 0.05, 0.1, and 0.5 probability, corresponding to confidence
intervals of 99\%, 95\%, 90\% and 50\%, are plotted in Figure
\ref{fig.results}a.  The location of maximum K-S probability is marked
as an ``x'', but we caution against overinterpretation of this
location because even the contour of lowest confidence (50\%) is very
broad.

If the detection efficiency is near unity for these stars, agreement
between {\it Kepler} and Doppler observations favors a high $\alpha$
but also demands implausibly high values of $\sigma_0$.  Better
reconciliation between {\it Kepler} and Doppler can be achieved if
{\it Kepler} detections are incomplete relative to the idealized
calculations for these stars, i.e. if $C < 1$.  If $C = 0.5$, then
$\sigma_0 \sim 2$ \mps~permits values of $\alpha \sim 4$, but not much
lower values: values of $\alpha \le 2$ are possible only if $\sigma_0
> 3.2$~\mps (total systematic noise $> 4.8$~\mps) at 99\% confidence
(Figure \ref{fig.results}b).

One cause of $C < 1$ may be interloping giant stars in the target list
\citep{Basri2011}.  Giant stars have radii $\ge 10R_{\odot}$, and the
transits of planets even as large as Neptunes will be $\le$ 12~ppm and
undetectable by {\it Kepler}, especially with the confounding effect
of oscillations \citep{Huber2010a}.  There is indirect evidence for
such contamination in the distribution of planet candidates with
stellar colors.  Figure \ref{fig.grjk} plots the $g-r$ (SDSS) and
$J-K$ (2MASS) colors of {\it Kepler} target stars from the KIC
\citep{Brown2011}.  Yellow and red points have estimated surface
gravities $4 < \log g < 4.9$ (putative main sequence stars) and $\log
g < 4$ (putative subgiants and giants), respectively.  Black contours
are lines of constant (sub)giant fraction.  Purple points mark
candidate planet hosts.  The green contour encloses 90\% of stars with
$T_e$=3660-4660~K.  Planet-hosting stars are conspicuously sparse in
the vicinity of $J-K \approx 0.7$ and $g-r \approx 0.9$, where the
fraction of (sub)giants exceeds 50\%.  Many putative K dwarf stars in
this region of color space may instead be misclassified (sub)giants,
with much larger radii and higher variability.

We also evaluated the range of ($\sigma_0$,$\alpha$) parameter space
over which the specific scenarios of rock-metal planets ($\alpha =
3.85$) and gas/ice-rich planets ($\alpha = 2$) are allowed (Figures
\ref{fig.results}c and d).  The former is permitted by a plausible
range of $\sigma_0$ for $C < 1$, with $C$ = 0.4-0.5 being most
consistent with our Doppler data.  All cases with $\alpha = 2$ are
ruled out at $>$95\% confidence as long as $C > 0.2$ and $\sigma_0 \le
2.4$~\mps (total systematic noise $\le 3.8$~\mps).

Small planets may instead comprise an admixture of rocky, ice-rich,
and gas-rich worlds.  \citet{Wolfgang2011} find evidence for a mixed
population around solar-type stars.  We considered this scenario by
assuming that the population consists of a mixture of $\alpha = 3.85$
and $\alpha=2$ planets with frequency $\gamma$ and $1-\gamma$,
respectively.  The K-S probability distribution vs. $\sigma_0$ and
$\gamma$ is plotted in Figure \ref{fig.gamma}.  The maximum K-S
probability (93\%) occurs for $\gamma = 0.88$ and $\sigma_0 =
2.8$~\mps, but a range of correlated $\gamma$ and $\sigma_0$ values
are possible.  If $\sigma_0$ is not much larger than 2~\mps~then
values of $\gamma$ near unity are clearly favored, and if $\sigma_0 <
2.6$~\mps~($<$4~\mps~total systematic Doppler error) then $\gamma >
0.5$ at 99\% confidence.

\section{Sensitivity to Model Assumptions \label{sec.tests}}

We performed a series of calculations to test the sensitivity of our
results to some assumptions of the model.  We considered the $C = 0.5$
case, and thus outcomes should be compared to Figure
\ref{fig.results}b.

{\it Distribution of jitter RMS:} We replaced the Rayleigh
distribution of jitter RMS $\sigma_0$ with an exponential
distribution, while maintaining the same ensemble RMS.  This
modification shifts the locus of acceptable models to values slightly
lower values of $\alpha$ and sightly higher values of $\sigma_0$
(Figure \ref{fig.tests}a), but otherwise does not significantly impact
our results.

{\it Correlated noise:} Correlated or ``red'' instrument noise in
Doppler observations does not decrease as the square root of the
number of measurements, making {\it de novo} detections of signals
comparable to such noise very difficult.  This has little impact on
our results because they rely on {\it Kepler} for planet detections
and we analyze only the variance (total power) of the RV, a quantity
independent of the noise spectrum.  Correlated noise would only be
important if there was significant drift of HIRES measurements on
timescales longer than the timespan of our measurements (months to
years).  Long-term monitoring of RV-stable stars rules out such
behavior, e.g. \citet{Apps2010}.  We further tested the possible
effect of red noise on our analysis by modeling instrument noise as
correlated with a power spectrum $exp(-\omega \tau)$, where $\tau$ is
the noise coherence time.  Uncorrelated (``white'') noise values $w_i$
at times $t_i$ are replaced by ``red'' noise values $r_i$ where
\begin{equation}
r_i = \sum_j \frac{w_j}{1 +\left[2\left(t_i - t_j\right)/\tau\right]^2},
\end{equation}
and the sum is over {\it all} observations, whether they are of a
given star, or not.  In calculating reddened instrumental noise, we
use the actual epochs of the observations $t_i$.  Errors are then
re-normalized to keep the variance the same.  The coherence time of
HIRES instrument noise is not known but we assume $\tau = 20$~d.
Figure \ref{fig.tests}b shows that the impact on our results is very
small.

{\it Random errors in KIC radii:} Inferences about densities and a
mass-radius relationship depend sensitively on {\it Kepler}'s
estimates of planet radii, which are uncertain.  To investigate the
effect of random errors, we added gaussian-distributed errors with
25\% RMS to the {\it Kepler} radii.  This modification broadens the
locus of acceptable parameter values and shifts the best-fit models to
slightly lower $\alpha$ and slightly higher jitter, but otherwise does
not significantly impact our results (Figure \ref{fig.tests}c).

{\it Systematic errors in KIC radii:} The astroseismically-determined
radii of many {\it Kepler} solar-type stars are systematically larger
(a median of 20\%) than KIC estimates \citep{Verner2011}.  If this
were also the case for the late K and early M stars in our sample, the
planets they host would be larger by the same amount, and hence less
dense.  If the effect is uniform, the inferred frequency of planets,
which depends mostly on detectability, transit depth and hence the
ratio of radii, is largely unchanged.  We investigated this scenario
by increasing the radii of all stars and planets by 20\% (Figure
\ref{fig.tests}d).  Larger planet radii and lower densities shift the
locus of permissable $\alpha$ and $\sigma_0$ to only slightly lower
values.  On the other hand, \citet{Muirhead2011} point out that a
stellar evolution model predicts consistently {\it smaller} radii for
planet-hosting M dwarfs compared to KIC estimates.  A running median
of KIC radii vs. effective temperature, reduced by 15\%, is roughly
consistent with a Yale-Yonsei 5~Gyr solar-metallicity isochrone.  We
therefore performed a second analysis in which star and planet radii
were uniformly decreased by 15\% (Figure \ref{fig.tests}d).  As
expected, this shifts the locus to both higher $\alpha$ and
$\sigma_0$.  As we discuss below, systematic overestimation of stellar
radius and the presence of interloping giant stars may not necessarily
be incompatible.

\section{Discussion \label{sec.discussion}}

Our combined analysis of {\it Kepler} transit detections and Doppler
radial velocities for late K and early M stars finds that consistency
is possible for a wide but not unlimited range of parameters.  As
expected for an analysis based on RV variance, there is an inverse
relationship between acceptable values of planet mass, i.e., the
power-law index $\alpha$ of the planet mass-radius relation, and
stellar jitter, i.e. the parameter $\sigma_0$ that characterizes its
distribution among stars.  However, if the level of radial velocity
jitter in M2K stars is as expected, reconciliation of {\it Kepler} and
Doppler observations can only be achieved if $\alpha \sim 4$, and
$\alpha \sim 2$ is excluded. In other words, small planets around
these stars are primarily rocky-metal ``super-Earths'' rather than
hydrogen gas-rich ``mini-Neptunes''.  We cannot absolutely rule out
higher jitter ($\sigma_0 \ge 3$~\mps, corresponding to total
systematic RMS $> 4.5$~\mps) that would admit a lower value of
$\alpha$, but there is no evidence to support such a choice.  Instead,
$\sigma_0 \sim 2$~\mps~is supported by the RMS of our paired Doppler
observations, the predicted stellar jitter based on chromospheric
activity and the observed levels of jitter among other, similar stars
\citep{Apps2010,Isaacson2010a}.  Our choice of $\alpha = 2$ to
represent gas-rich planets is conservative because theoretical
modeling suggests values closer to zero or even negative over the mass
range of interest \citep{Rogers2011}.

Reconciliation of {\it Kepler} and Doppler data, even with $\alpha
\sim 4$, also appears to require that {\it Kepler}'s detection
efficiency be less than unity and perhaps $\sim$50\%.  Some of this
incompleteness could arise if many target stars are misclassified
subgiant or giant stars around which Neptune-size or smaller planets
are difficult or impossible to detect by {\it Kepler}.  Spectroscopic
follow-up finds that essentially all late K and M {\it Kepler} stars
brighter than $K_p = 14$ are giants (Mann et al., in prep.); we
estimate the rate of interlopers in our sample of {\it Kepler} targets
to be at least 15\%.  Giant interlopers are rare among the transiting
planet-hosting {\it Kepler} stars \citep{Muirhead2011} because the
vast majority of planets are smaller than Jupiter and not detectable
around giant stars.  Additional incompleteness could come from higher
stellar variability.

Our analysis appears robust to the precise choice of function for the
distribution of jitter RMS among stars, as long as the overall noise
variance is conserved.  It is also insensitive to the presence of
correlated or ``red'' noise in the Doppler RV data.  Although our
results are not overly susceptible to random errors in estimated
stellar radii, they do vary with uniform systematic errors in those
values.  If radii have been uniformly overestimated, as comparisons
with stellar evolution models suggest, agreement between {\it Kepler}
and M2K statistics favors a slightly higher value of $\alpha$,
reinforcing our conclusion that the small planets around these stars
are primarily rocky.  Although the resulting offset of the locus with
$\alpha$ may seem small, one property of a power-law MRR is that a
compensatory fractional change in index $\alpha$ will equal the
fractional magnitude of a systematic change in radius, modulo a
logarithmic factor which is approximately unity.  For example, if
radii are 15\% smaller then $\alpha$ should be 4.6 instead of 4.

Systematic underestimation of stellar radii can be reconciled with the
presence of interloping giant stars by accounting for strong selection
effects among stars with transit-detected planets: Just as transit
surveys of a given set of stars are biased towards the largest planets
\citep{Gaudi2005}, a given set of planets will be more readily
detected by transit around the smallest stars in a sample; stars with
detected planets are thus not necessarily representative of the entire
sample.  Reliable estimates of the radii of a presentative sample of
late-type {\it Kepler} target stars should be vigorously pursued.
    
\begin{figure}
\epsscale{0.7}
\plotone{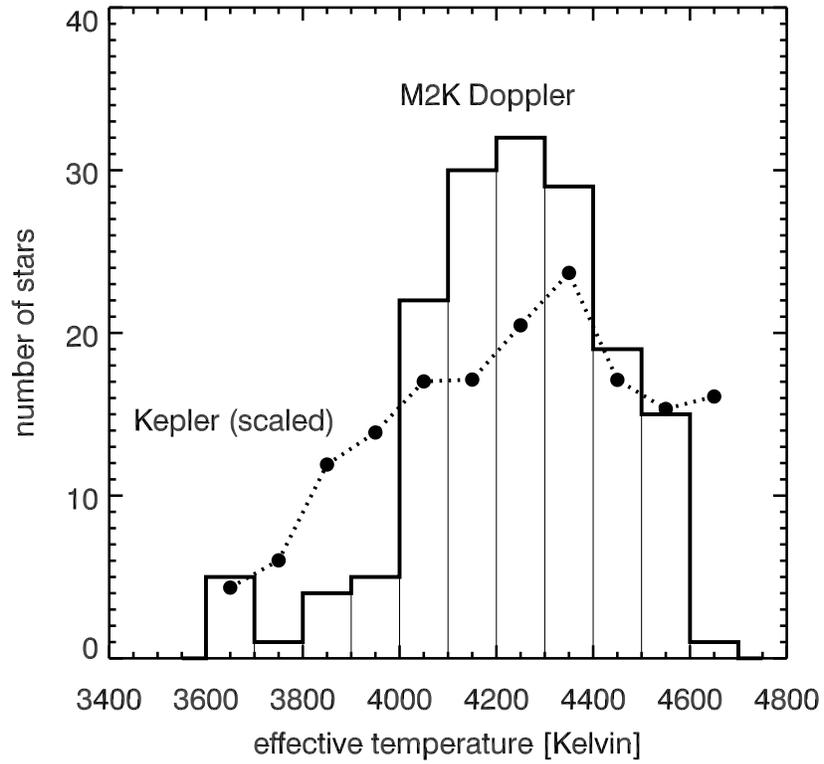}
\caption{Distribution of effective temperatures in the M2K Doppler
  survey (bars) and {\it Kepler} Quarter 2 target catalog (dotted
  line) in the range 3660-4660~K.  This range was chosen to maximize
  the similarity between the distributions as measured by the
  Kolmogorov-Smirnov test.\label{fig.teff}}
\end{figure}

\begin{figure}
\epsscale{0.7}
\plotone{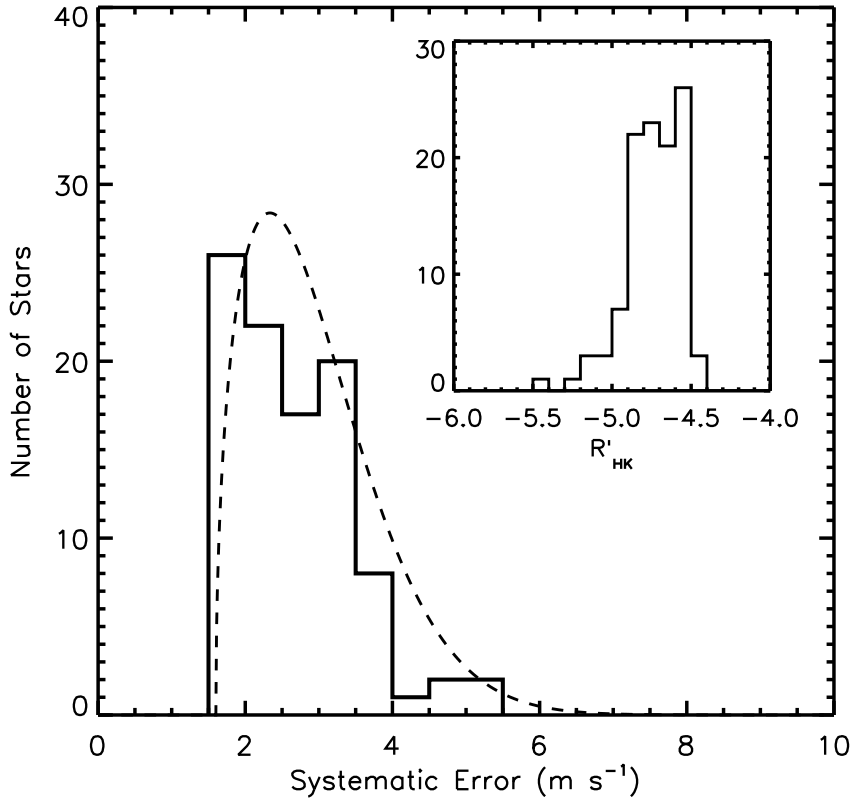}
\caption{Predicted systematic error (instrument and stellar jitter) of
  Doppler measurements for 100 M2K stars based on $B-V$ color and
  emission in the H and K lines of Ca II.  The dashed line is a
  best-fit noise model with a uniform instrument error of 1.6 \mps~RMS
  added in quadrature to Rayleigh-distributed stellar jitter with
  $\sigma_0 = 1.7$ \mps.  The distribution of Ca II HK emission,
  parameterized by the $R'_{HK}$ index, is plotted in the inset.  The
  median $R'_{HK}$ of the sample is -4.70, corresponding to an age of
  about 3~Gyr \citep{Mamajek2008}.  \label{fig.activity}}
\end{figure}

\begin{figure}
\epsscale{0.8}
\plotone{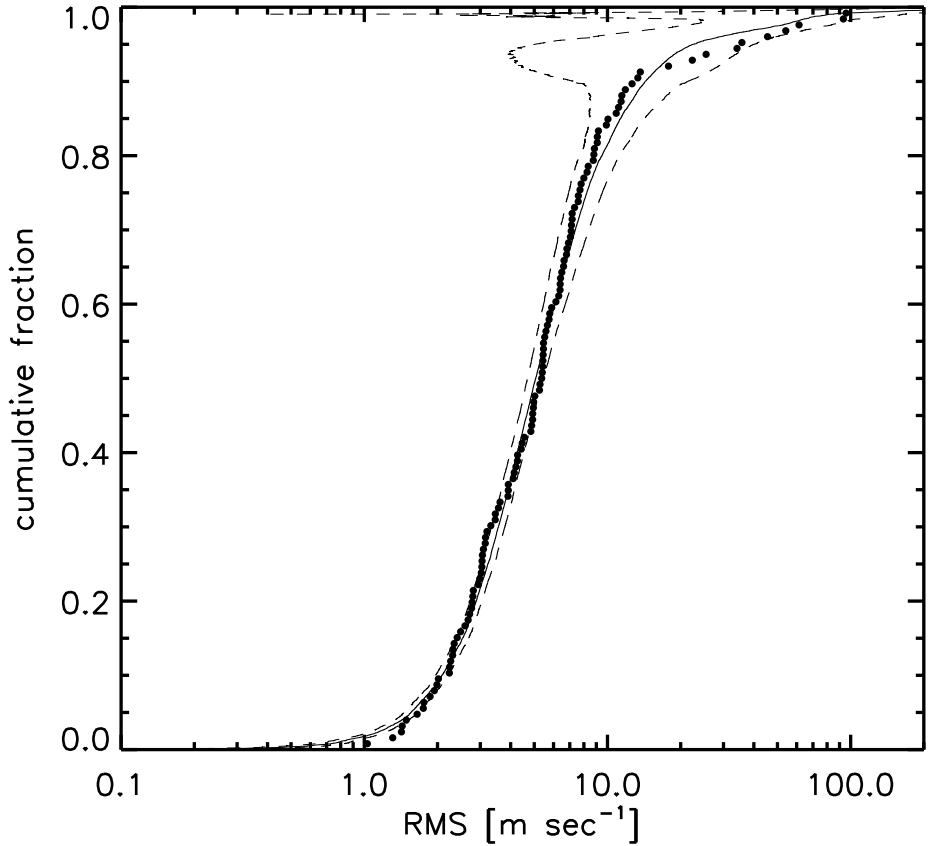}
\caption{Cumulative distribution of RV RMS (points) in M2K stars with
  $T_e = 3660-4660$~K.  The solid line is a model based on {\it
    Kepler} radii with Rayleigh-distributed systematic noise
  ($\sigma_0=2.6$~\mps), {\it Kepler} detection efficiency factor $C =
  0.5$, binary cut-off $B=110$~\mps, and power-law mass-radius
  relation with index $\alpha=3.85$ for planets with $R_p \le
  3$\rearth.  The observations and model have a two-sided
  Kolmogorov-Smirnov probability of 93\% that they could be drawn from
  the same population.  The dashed lines are 95\% confidence intervals
  for uncertainties generated by the finite size of the {\it Kepler}
  sample.  These intervals are illustrative; cumulative distributions
  do not reverse.  \label{fig.rmsdist}}
\end{figure}

\begin{figure}
\epsscale{1.0}
\plotone{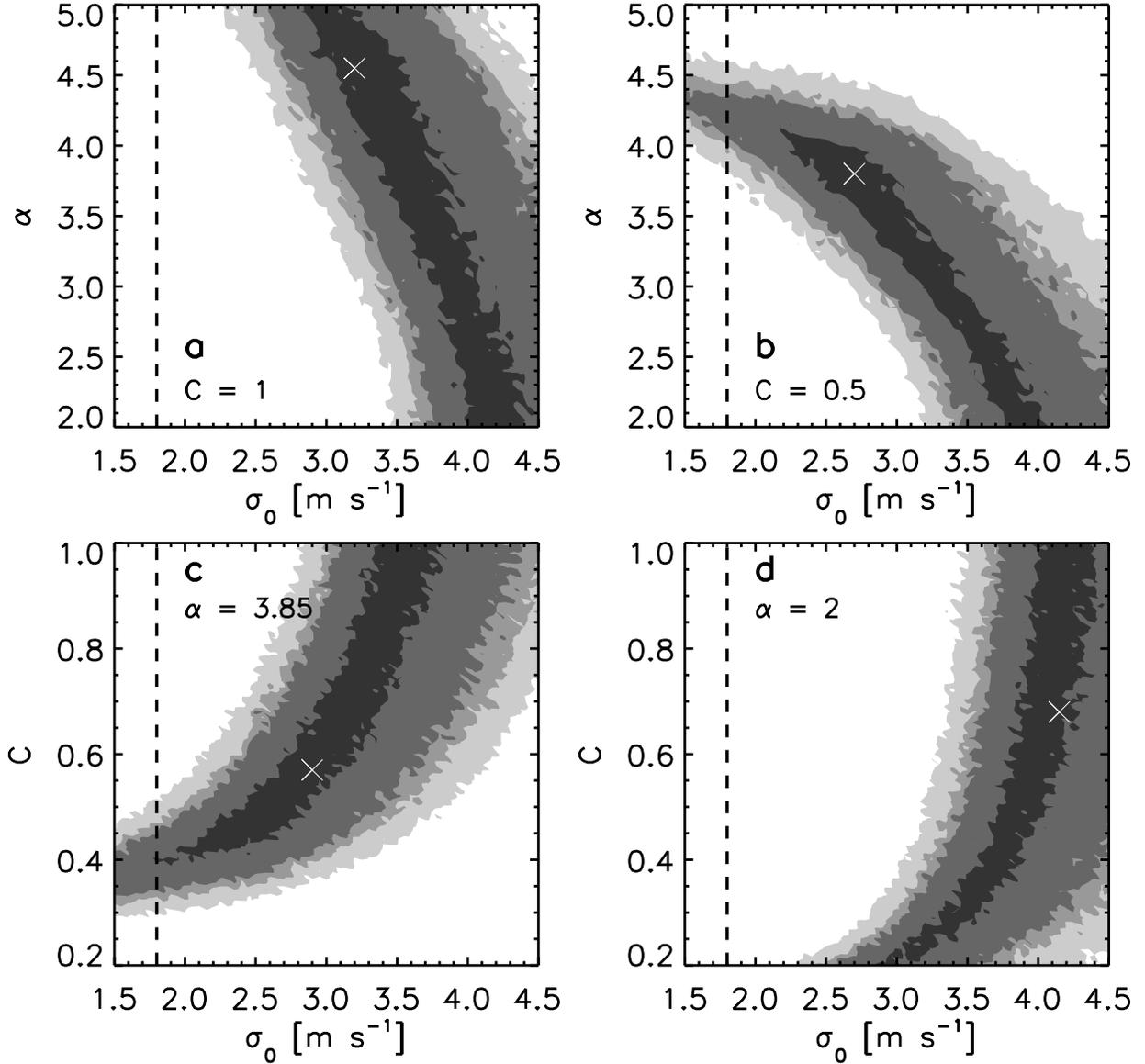}
\caption{Kolmogorov-Smirnov probability for our {\it Kepler}/M2K
  comparison as it varies with jitter parameter $\sigma_0$,
  mass-radius relation power law index $\alpha$, and {\it Kepler}
  detection efficiency $C$.  Contours (lightest to darkest) are K-S
  probabilities of 0.01, 0.05, 0.1, and 0.5, representing confidence
  intervals of 99\%, 95\%, 90\%, and 50\%.  The X marks the location
  of maximum probability, and the vertical dashed line marks $\sigma_0
  = 1.8$~\mps, the value derived from observations and expected based
  on the distribution of Ca II HK emission among M2K stars.  Panel (a)
  plots K-S probabilities with $\sigma_0$ and $\alpha$ assuming a {\it
    Kepler} detection effiency of $C = 1$; (b) same as (a) but with $C
  = 0.5$; (c) distribution with $\sigma_0$ and $C$ for a rocky planet
  MRR ($\alpha=3.85$); (d) same as (c) except for a notional gas-rich
  planet MRR ($\alpha=2$). \label{fig.results} }
\end{figure}

\begin{figure}
\epsscale{0.8}
\plotone{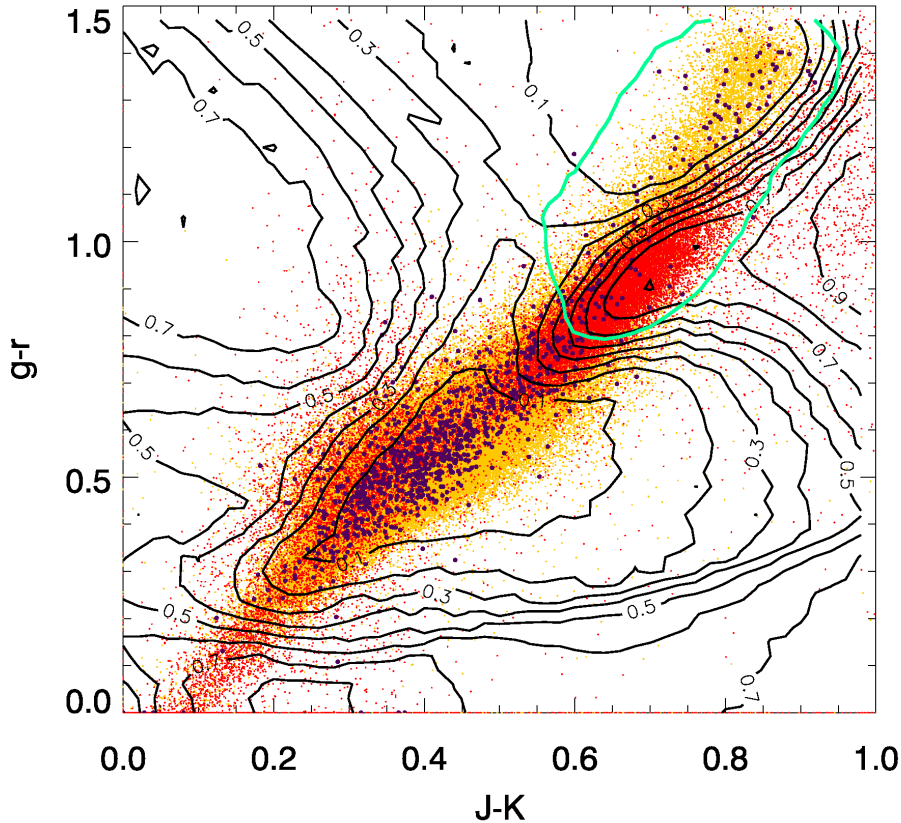}
\caption{Color-color (SDSS $g-r$ and 2MASS $J-K$) diagram of {\it
    Kepler} target stars.  Yellow and red points represent stars with
  estimated $\log g > 4$ (putative dwarfs) and $\log g < 4$ (putative
  giants or subgiants), respectively.  Black contours are of constant
  (sub)giant fraction and the green contour encircles 90\% of stars
  with $T_e=3660-4660$~K.  The large purple points are the host stars
  of planet candidates.  The main sequence and giant branches
  intersect in the region of color-color space occupied by late K
  stars.  This region appears to be deficient in planet candidates and
  those present are found where the (sub)giant fraction is least.
  This suggests that many putative dwarf stars in this region may be
  misclassified subgiants or giants, around which planets would be
  more difficult or impossible for {\it Kepler} to
  detect. \label{fig.grjk} }
\end{figure}

\begin{figure}
\epsscale{0.7}
\plotone{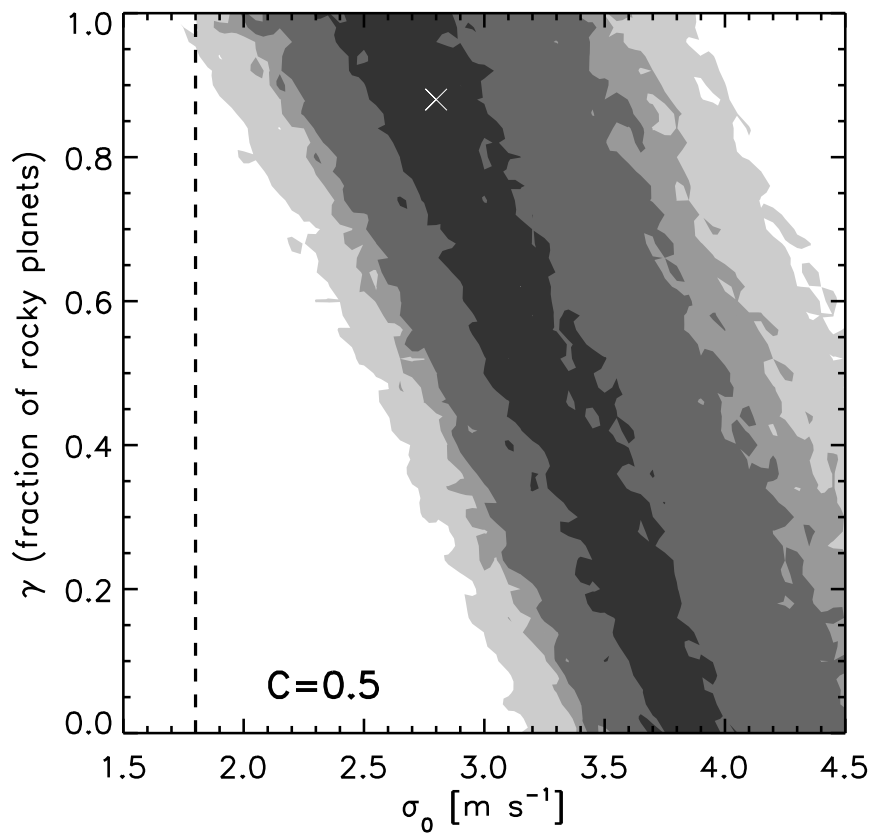}
\caption{Kolmogorov-Smirnov probabilities vs. jitter magnitude
  $\sigma_0$ and fraction $\gamma$ of rocky ($\alpha=3.85$) planets
  vs. gas-rich ($\alpha=2$) planets, assuming a {\it Kepler}
  completeness $C=0.5$.  Countours (lightest to darkest) are K-S
  probabilities of 0.01, 0.05, 0.1, and 0.5, representing confidence
  intervals of 99\%, 95\%, 90\%, and 50\%.  The X marks the location
  of maximum probability.  The vertical dashed line at 1.8~\mps~marks
  the expected value of $\sigma_0$ for M2K stars. \label{fig.gamma} }
\end{figure}

\begin{figure}
\epsscale{1.0}
\plotone{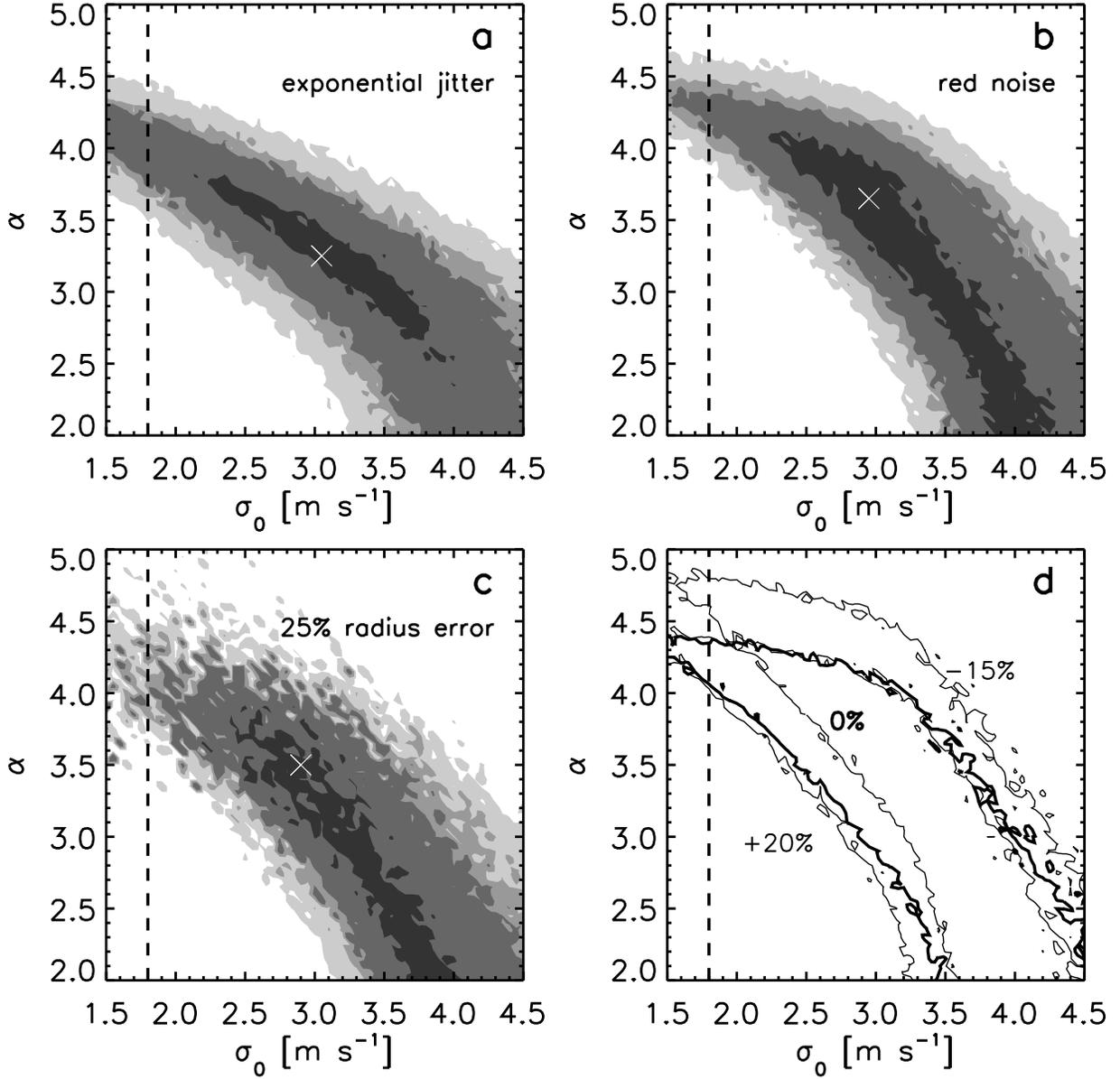}
\caption{Sensitivity of our results to different assumptions in the
  model used to translate {\it Kepler} radii into Doppler radial
  velocity variance.  All calculations assume $C = 0.5$ and these
  plots should be compared to Figure \ref{fig.results}b.  In (a) we
  use an exponential rather than a Rayleigh function to describe the
  distribution of jitter RMS among stars.  In (b) instrument noise is
  modeled as being correlated on a timescale of 20~d.  In (c) 25\%
  gaussian error is added to {\it Kepler} stellar radius estimates.
  In (d) the radii of {\it Keper} stars (and planets) are uniformly
  increased by 20\% or decreased by 15\% from KIC values (heavy line).
  See text for justification of these choices.  For clarity, only the
  90\% confidence contours are shown in the last
  panel. \label{fig.tests}}
\end{figure}

Our analysis is predicated on statistically indistinguishable planet
populations in our samples of stars from the {\it Kepler} field and
solar neighborhood.  A plausible condition for this assumption is that
the two samples have similar mass and metallicity distributions and be
drawn from the same stellar population.  The effective temperature
distributions are similar, but not identical (Figure \ref{fig.teff})
and this may translate into differences in stellar mass.  There is an
excess of about 30 M2K stars ($\sim$20\% of the sample) around 4200~K
and a deficit around 3850~K.  According to a Yale-Yonsei 5~Gyr
solar-metallicity isochrone, this 350~K increase corresponds to
changing the stellar mass from 0.57\msun~to 0.65\msun.  Adopting the
relation between stellar mass and giant planet frequency of
\citet{Johnson2010} at face value, these M2K stars would have a 14\%
higher incidence of giant planets, but the giant planet frequency in
the overall sample would only be 3\% higher.  According to Equation 9
in \citet{Howard2011} the frequency of {\it all} planets would
decrease by 6\%.

M2K stars are all within 45~pc of the Sun, and the median distance is
25~pc, placing them well within the galactic disk.  These stars are
drawn from a proper motion-selected catalog ($>$40~mas~yr$^{-1}$) with
a transverse velocity limit of 8.6~km~s$^{-1}$ at 45~pc
\citep{Lepine2005}.  The velocity dispersion of stars in the solar
neighborhood is anisotropic but a rough estimate of 80\% completeness
at 45 pc is obtained by assuming an isotropic distribution with a
dispersion of 25~km~s$^{-1}$ \citep{Bond2010}.  The correlation
between metallicity and velocity dispersion (via age) means that this
sample will be biased against metal-rich stars, but this effect is
very small: Stars with [Fe/H]=-0.5 (more metal-poor stars are very
uncommon) have a velocity dispersion $\sim$5~km~s$^{-1}$ higher than
their solar metallicity counterparts \citep{Lee2011}, and the
corresponding completeness is $\sim$84\% at 45~pc.  The bias against
solar-metallicity stars in M2K is therefore $\le$5\%.  Although we
excluded the most active stars from the analysis and may have removed
any very young stars, this should not affect the metallicity
distribution because the metallicity-age relation is flat in this
range \citep{Holmberg2007}.  Tidal decay of the orbits of low mass
planets around small stars is expected to be extremely slow
\citep{Jackson2009} and would not appreciably evolve a planet
population.

The kinematics and metallicities of {\it Kepler} field stars have yet
to be established. The center of the {\it Kepler} field
($l=77^{\circ}$, $b=+13^{\circ}$) is nearly perpendicular to the
direction to the galactic center, and approximately parallel to the
galactic plane.  We estimate photometric distances from {\it Kepler}
derived temperatures using the empirical relation for absolute
magnitude $M_J \approx 6.25 - 16.53 \log (T_e/4000)$.  At the median
estimated distance of the subsample (256~pc), a star at the center of
the {\it Kepler field} has approximately the same galactocentric
distance as the solar neighborhood and is only $\sim$60~pc above the
galactic plane; most stars should belong to the thin disk and have
near-solar metallicities.  Consistent with this, the TRILEGAL stellar
population model \citep{Vanhollebeke2009} predicts that only 7\% of
stars in this range of effective temperature and magnitude belong to
the thick disk or halo, and only 6\% have [Fe/H] $< -0.5$.  Thus we
conclude that the {\it Kepler} and M2K samples are very similar in
mass, metallicity, and age.

\citet{Howard2011} estimated planet densities by comparing
distributions of {\it Kepler} radii with masses from a Doppler survey
of solar-type stars.  They inferred a higher density for the smallest
planets, consistent with our findings.  \citet{Wolfgang2011}, using a
different set of Doppler-detected planets, also concluded that the
majority of small planets around solar-type stars are rocky.  They
also found that the proportion of low density, gas-rich planets
increases with planet size, a feature essentially intrinsic to our
analysis because of our choice of $\alpha = 2$.  Theoretical models
predict the formation of inner, rocky planets \citep{Raymond2004}, and
the stellar UV-driven escape of any primordial hydrogen atmospheres
\citep{Pierrehumbert2011a}.  The low density of the short-period
super-Earth GJ~1214b can be explained by a substantial hydrogen
envelope \citep{Charbonneau2009a,Croll2011}, but also by a thick
H$_2$O shell \citep{Bean2010a,Desert2011}.  Its host is a cooler
(3000~K), much less luminous mid-M star and this may permit retention
of hydrogen \citep{Pierrehumbert2011a}.  Gas- and ice-rich planets
resembling GJ~1214b may be the exception rather than the rule around
the coolest {\it Kepler} target stars.  Refinement of Doppler
systematic errors and the properties of {\it Kepler} target stars,
specifically the radii of K and M dwarfs and the fraction of
interloping giants, will permit more robust constraints.

\acknowledgments

This research was supported by NSF grants AST-09-08406 (EG),
AST-10-36283 (DAF), AST-09-08419 (SL); NASA grants NNX10AI90G and
NNX11AC33G (EG), and the NASA KPDA program (DAF).  The {\it Kepler}
mission is funded by the NASA Science Mission Directorate, and data
were obtained from the Multimission Archive at the Space Telescope
Science Institute, funded by NASA grant NNX09AF08G.  We thank Andrew
Howard for his help with screening Doppler template spectra.

\clearpage

\clearpage

\begin{deluxetable}{ccrr}
\tablecaption{Radial velocity statistics of 150 stars with $T_e$=3660-4660~K in the M2K Survey \label{tab.obs}}
\tablewidth{0pt}
\tablehead{
\colhead{Star} & \colhead{Measurements} & \colhead{Stand. Dev.} & \colhead{Formal Error}\\ 
&  & \colhead{m s$^{-1}$} & \colhead{m s$^{-1}$}
}
\startdata
1 & 18 & 3.19 & 1.30 \\
2 & 8 & 5.42 & 1.58 \\
3 & 9 & 2.94 & 1.07 \\
\enddata
\tablecomments{Table \ref{tab.obs} is published in its entirety in the 
electronic edition of the {\it Astrophysical Journal}.}
\end{deluxetable}

\end{document}